# Case Study-Based Approach of Quantum Machine Learning in Cybersecurity: Quantum Support Vector Machine for Malware Classification and Protection

Mst Shapna Akter
*Dept. of Computer Science*
*Kennesaw State University*
Kennesaw, USA
Email: makter2@students.kennesaw.edu

Hossain Shahriar
*Dept. of Information Technology*
*Kennesaw State University*
Kennesaw, USA
Email: hshahria@kennesaw.edu

Sheikh Iqbal Ahamed
*Dept. of Computer Science*
*Marquette University*
Wisconsin, USA
Email: sheikh.ahamed@marquette.edu

Kishor Datta Gupta
*Dept. of Computer Science*
*Clark Atlanta University*
Georgia, USA
Email: kgupta@cau.edu

Muhammad Rahman
*Dept. of Information Technology*
*Clayton State University*
Morrow, USA
Email: muhammadrahman@clayton.edu

Atef Mohamed
*Dept. of Information Technology*
*Georgia Southern University*
Georgia, Country
Email: amohamed@georgiasouthern.edu

Mohammad Rahman
*Dept. of Electrical and Computer Engineering*
*Florida International University*
Florida, USA
Email: marahman@fiu.edu

Akond Rahman
*Dept. of Computer Science and Software Engineering*
*Auburn University*
Auburn, USA
Email: akond@auburn.edu

Fan Wu
*Dept of Computer Science*
*Tuskegee University*
Tuskegee, USA
Email: fwu@tuskegee.edu

*Abstract*—Quantum machine learning (QML) is an emerging field of research that leverages quantum computing to improve the classical machine learning approach to solve complex real-world problems. QML has the potential to address cybersecurity-related challenges. Considering the novelty and complex architecture of QML, resources are not yet explicitly available that can pave cybersecurity learners to instill efficient knowledge of this emerging technology. In this research, we design and develop QML-based ten learning modules covering various cybersecurity topics by adopting student centering case-study based learning approach. We apply one subtopic of QML on a cybersecurity topic comprised of pre-lab, lab, and post-lab activities towards providing learners with hands-on QML experiences in solving real-world security problems. In order to engage and motivate students in a learning environment that encourages all students to learn, pre-lab offers a brief introduction to both the QML subtopic and cybersecurity problem. In this paper, we utilize quantum support vector machine (QSVM) for malware classification and protection where we use open source Pennylane QML framework on the drebin215 dataset. We demonstrate our QSVM model and achieve an accuracy of 95% in malware classification and protection. We will develop all the modules and introduce them to the cybersecurity community in the coming days.

*Index Terms*—Quantum Machine Learning (QML), Quantum Support Vector Machine (QSVM), Cybersecurity, Malware Classification

## I. INTRODUCTION

Case study based learning is a well-adopted learning approach in education and refers to learner-centered strategies and features intense interaction between participants as they increase their knowledge and analyze the case as a group. Moreover, students are involved in discussion of certain scenarios that approximate or are typically real-world examples [1]. Towards providing students with the crucial skills they need through hands-on experiences in solving real-world security problems, case study based learning provides a unique hands-on approach comprising pre-lab, lab, and post-lab activities [2]. Among those three activities, pre-lab demonstrates elementary concepts on a topic, which will help students gather prior knowledge required for doing experimental work. The next step comes with the lab, which provides the necessary instruction of hands-on practice. This step engages and inspires students to solve real-life problems with optimal solutions and enhance self-confidence through mastery-building experience. The post-lab is an extension of hands-on lab practice, mainly focusing on optimizing and developing further solutions. This one is crucial for growing students' self-efficacy through peer performance observation and creativity development. The following steps are included in the case-study based learning model [ Figure 1 shows the graphical representations ]

Step 1: Initiating/ understanding through pre-lab instructions.
Step 2: Engaging/analyzing problems with the hands-on lab on real-life issues.
Step 3: Optimize the solutions of the hands-on lab using different approaches.
Step 4: Repeat steps 1-3 on various algorithms or datasets.

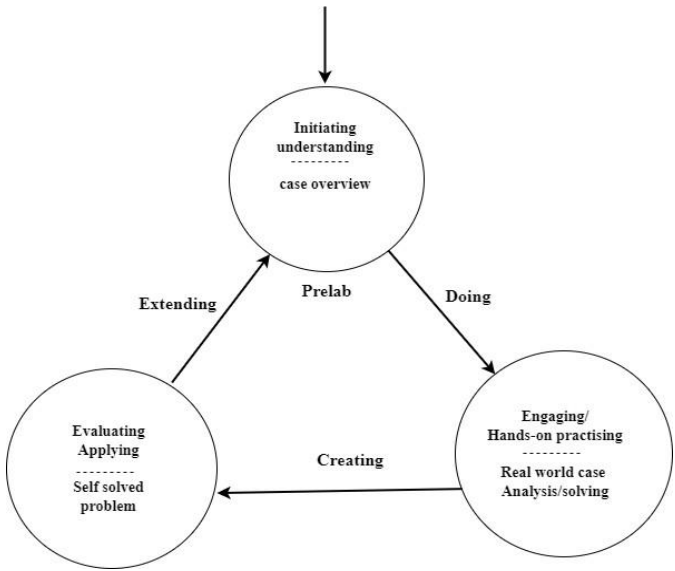

Fig. 1: Case-study based Learning Model

Recently, case study based learning emerged with massive popularity in the field of Cybersecurity [3, 4]. For instance, Kam et al. [5] used a learning platform using web based Learning Content Management Systems (LCMS) and a mobile client application (app) for raising students' knowledge of basic Cybersecurity and privacy issues. They developed a web-based technology called "ComicBEE" which includes "choose-your-own-adventure". The features allows readers to understand cyber security through characters and storyline illustrations [5]. Similarly, Giannakas et al. [6] applied case study based learning approach for developing several projects on Cyber hacking field.

Many institutions offer Quantum machine learning and cybersecurity-related courses in their core course curriculum. Still, those do not seem to provide sufficient learning materials, including open-source portable hands-on labware systems, especially dealing with malware protections. Since the hands-on lab comes with many challenges, including: the scarcity of knowledgeable instructors, configuration of the whole methods; a collection of required resources and materials; and finally, dedication to finish all steps. We tried to overcome the difficulties and made available our open-source, portable, modular, and easy-to-adopt QML approaches in the field of cybersecurity. The approaches can be found online, which are structured as modules with three parts - Pre-lab, hands-on lab, and post lab. The 10 learning modules are: M0: Getting Started with CoLab on ML for CyberSecurity; M1: Malware Classification Using Quantum Support Vector Machine; M2: Malware Classification using Quantum Neural Network (QNN); M3: Intrusion Detection using Quantum Transfer Learning; M4: User Behavior Anomaly Detection using Quantum Variational Network; M5: Malicious Web Application Detection using Quantum Support Vector Machine; M6: Network Intrusion Detection using Quantum Neural Network; M7: Network Intrusion Detection Using Quantum Neural Network; M8: JavaScript Vulnerability Detection Using quantum Neural Network; M9: Malicious PDF Detection using quantum support vector machine; and M10: Android Malware Detection Using quantum Support Vector Machine.

In this paper, we present one of the modules covering overview on malicious link, which will lead to malware classification, Overview of Quantum Support Vector machine (QSVM), QSVM for malicious link (Getting started), QSVM for malware classification and protection (Getting started).

According to the US Cybersecurity and infrastructure security agency, Malware is a collective term for malicious programming codes, scripts, active content, or intrusive software that includes computer viruses, worms, ransomware, rootkits, trojan horses, dialers, adware, spyware, keyloggers, malicious Browser Helper Objects, and other forms that are intended to damage computer systems, programs, or mobile and web applications (BHOs) [7]. Malware can be characterized based on its intended use and means of dissemination and can include any software created with malicious intent. If a malicious application or file is installed on or shared with a new computer, the virus will automatically copy itself into the new computer and run its code [8]. Malware programs have the ability to self-execute, infecting a computer device without the user's knowledge or consent. Such dangerous files or programs originate from different sources, including the internet in general, file downloads from nefarious websites, or specifically clicking on a malicious link.

## II. RELATED WORK

Literature review shows the crucial development of case study-based and project-based learning approaches which highlight solving critical challenges related to countless disciplines because of their several benefits in terms of implementation.

For example, a case-study-based learning was implemented to provide a productive platform for conducting machine learning-based hands-on-lab exercises for cybersecurity trainees [9]. The authors presented a knowledge graph in cybersecurity dominion using the following procedures: first depicted a preliminary visualized model of the knowledge graph allowing students to search for cyberattacking and other corresponding concepts, then created a customized graph according to the standard and progression of each student and finally built a personalized lab and conducted a case-study based experiment to explore the learning outcomes. An experiment utilizing CyberKG was evaluated in a security class at Arizona State University where selected graduate students completed three hands-on labs. This case-study-based approach was considered ideal not only to measure the effectiveness of the prototype virtual lab platforms but for being comprehensible and capable to make desired suggestions for the CyberKG system.

Another study has been performed implementing a case-study-based module to engage learners of the University of Illinois at Urbana-Champaign in real-world ethical dilemmas ingrained within cybersecurity dimensions to develop consciousness, collective problem-solving abilities, communication skills, and interpretative ethical reasoning expertise essential for future cybersecurity professionals and to overcome a challenge of instigating ethical perspective into an entirely technical field [10]. The step-by-step approach to performing the study incorporates involving the learners in a group dialogue that fosters critical thinking for analysis, then incorporating ethical frameworks into the dialogue to permit as many solutions as possible, and finally performing an evaluation survey to obtain the learner's viewpoints and determine the best headway for cybersecurity-related issues.

In the same context, another study advocated the effectiveness of both case study and conventional lecture-based approach to determine which one is more powerful for rigorous scientific research in the software engineering realm [11]. With the help of a set of predefined parameters and questionnaires derived from the SE domain the study used to create a bridge between academia and industry and forecast the true character of software development. The authors highlighted the use of case studies speculated from real-world industrial projects and analyzed students' perspectives to nurture essential learning goals such as understanding, expertise, software security and management, safety issues, communication proficiencies, problem-solving skills, and behavior required by the industry. Finally, the method conveyed that case study-based learning is more effective and score higher compared to lecture-based learning in terms of the learning parameters.

Similarly, a project-based learning framework was followed to facilitate undergraduate students for evaluating and investigating the directions of cyber-attacks on a networked cyber-physical system (CPS) [12]. The paper is contextualized by describing the benefits of utilizing PBL technique in boosting the overall student accomplishments towards the learning outcomes established in the cybersecurity domain. The authors proposed a design that provides greater accuracy with minimal laboratory costs accompanying possible directions for future amplification in this context.

Additionally, another research presented an innovative approach for designing and application of undergraduate engineering programs using applied machine learning technology integrated with a project-based learning approach [13]. Research results successfully illustrated the combination of PBL and ML approaches as an extremely effective, application-focused, and encouraging engineering tool that will allow users the ability to recognize, construct and resolve complex real-world engineering problems. The authors delineated three different components adopted at different stages and discussed the applicability of the proposed project highlighting self-directed education rather than self-education.

## III. MALWARE

Malware, also known as malicious software, is basically intrusive software developed by hackers or, in general, referred to as cybercriminals [14]. The purpose of developing such malicious software is to damage or destroy computer systems. Some examples are ransomware, trojan viruses, worms, spyware, and adware [15].

**Ransomware:** Ransomware is one kind of malicious software that tries to gain access to a system to gain sensitive information and encrypts the information to ensure blocking users from the system. Such circumstance helps the hacker to demand financial credentials. A phishing scam is used to trap users in the ransomware process. Some disguised links are basically regarded as ransomware, which user downloads without being aware of. Attackers then encrypt some specific data, which can be decrypted using a mathematical key. When someone pays, they open the data [16].

**Trojan Virus:** Trojan viruses also follow the rules of being downloaded by the users. Most of the time, it is altered as beneficial software, but after being downloaded, it gets access to the sensitive data and either deletes or blocks, or modifies, which can be very harmful to the device. It is not designed to self-replicate like other viruses and worms [17].

**Worms:** Worms are malicious software that spreads quickly to all devices connected to a network. Worms can spread without a host application, unlike viruses. Through a network connection or a downloaded file, a worm infects a machine, where it multiplies and spreads at an exponential rate. Like viruses, worms have the ability to damage a device's functionality, and damage data [18].

**Spyware:** Spyware is malicious software that works silently on a computer and transmits data to a remote user. Instead of only interfering with a device's performance, spyware can provide attackers remote access to critical information and target it. Hackers routinely use spyware to obtain financial or personal information. Keyloggers are a specific type of malware that monitors keystrokes and releases passwords and other sensitive information [19].

**Adware**: Adware is malicious software that monitors computer usage in order to display pertinent ads. Although it is not always destructive, Adware can occasionally disrupt a system. Adware has the ability to direct browsers to malicious websites that contain Trojan horses and spyware. A PC can become noticeably slower if it has a lot of adware. Because not all Adware is harmful, it is essential to have protection that regularly and intelligently monitors these apps [20].

IV. CASE STUDY LABWARE DESIGN

The portable labware is developed, designed and deployed on the open source environment "Google Colaboratory (co-Lab)". This environment allows users to access and share from anywhere and anytime without installation and maintenance hassles. Therefore, learners can interactively collaborate with other peers and practice and run all modules. Each of our case study-based portable hands-on labware modules is developed based on real-life examples of malware, ransomware, anomaly, intrusion, and vulnerabilities of codes. The structure of the modules consists of three components: Pre-lab for basic knowledge including getting started with a Hello World example, hands-on lab with the in-depth explanation of experiments, post-add lab with instructions for further optimizations

### A. Pre-lab for basic knowledge including getting started

Pre-Lab outlines QML solutions to cybersecurity, such as prevention and detection, and introduces a specific cybersecurity study scenario with the root of security threats, attack plans, and effects. Students can watch, observe, and gain perspective insight processing by observing a simplified "hello world" example for the cybersecurity case and corresponding QML solution. This prepares students with a specific cybersecurity case for conceptual understanding and getting started experience with QML solutions for such cybersecurity cases. By utilizing the machine learning technique, it aids students in developing a fundamental understanding of why these cybersecurity vulnerabilities need to be fixed. A screenshot of the Pre-lab for Module 1's Support Vector Machine algorithm for Malware classification and prevention is shown in Figure 2.

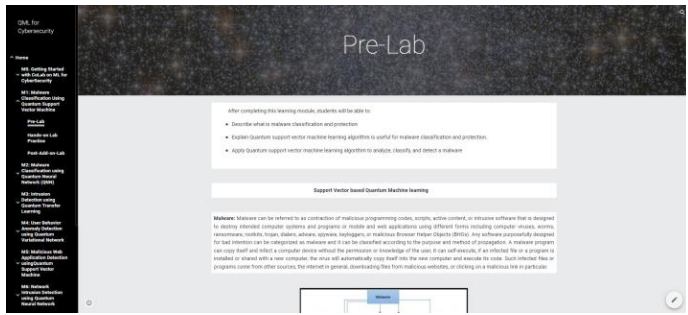

Fig. 2: Screenshot of Module 1 Pre-Lab

### B. Hands-on lab with the in-depth explanation of experiments

The open-source Google CoLab collaboration platform, which is an in-the-browser environment with a free Google cloud service, is used to design, create, and deploy the hands-on activity laboratories. To access CoLab and run the lab from any mobile device or laptop, students only need a Google account. After finishing the practical activity lab, students will have first-hand experience with problem-solving. Students can practice using the visual cues for direction more by using the screenshots for each stage. A screenshot of the lab from Module 1 is shown in Figure 3.

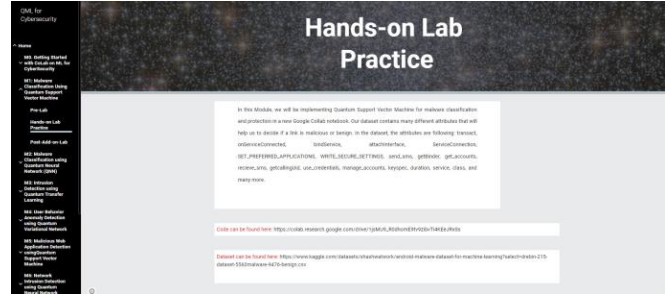

Fig. 3: Screenshot of Module 1 Hands on Lab

We use the drebin215 dataset used for Malware classification [21]. The dataset has 215 attributes. We used the following pipeline using Quantum Support Vector Machine ( QSVM ). After successfully loading the whole dataset, we click the 'Choose File' and upload the Malware dataset. Pennlylane library allows users to access the quantum computer and perform computational tasks. After downloading and importing all necessary libraries, we preprocessed the dataset. Finally, we fed the processed data into the SVM algorithm and performed the computation in the quantum circuit. We used the following pipeline using Quantum Support Vector Machine ( QSVM ). After successfully loading the whole dataset, we click the 'Choose File' and upload the Malware dataset. Pennlylane library allows users to access the quantum computer and perform computational tasks. After downloading and importing all necessary libraries, we preprocessed the dataset. Finally, we fed the processed data into the SVM algorithm and performed the computation in the quantum circuit. We evaluated our model with evaluation metrics such as accuracy [22], precision [23], recall [24], and f1 score.

### C. post-add lab with instructions for further optimizations

Students are encouraged to use their newly acquired knowledge and skills to address actual problems in the Post add-on lab. It encourages critical reflection on the provided example and practical application for improving problem-solving, such as raising the prediction and detection accuracy rate with new innovative concepts and active testing and experiments. With Colab, students can share their original work with others on the cloud.

V. STUDENT LEARNING ASSESSMENT

We conducted a preliminary survey collected from a total of sixteen undergraduate Engineering students at Kennesaw State University. Surveys are represented in quantitative and qualitative views. We conducted both a pretest and post-test survey, where we asked various questions.

**A. Pre-Test Survey:** Among sixteen students, all of them considered themselves in the age group between 18 and 25 years. We asked the participants to describe their level of education in the field (a) Quantum Machine Learning (b) cyber security. We further asked question about (c) preference on on project based lab work than by listening to lectures (d) preference on personally doing or working through examples (e) preference having a learning/tutorial system that provides feedback. Figure Figure 5, 6, 7, and 8 displays the responses

TABLE I: display the responses of students on age group

| # | Answer | % | Count |
|---|---|---|---|
| 1 | < 18 years | 0.00% | 0 |
| 2 | Between 18 and 25 years | 20.00% | 16 |
| 3 | Between 26 and 35 years | 80.00% | 0 |
| 4 | Between 36 and 45 years | 0.00% | 0 |
| 5 | Between 46 and 55 years | 0.00% | 0 |
| 6 | >55 years | 0.00% | 0 |

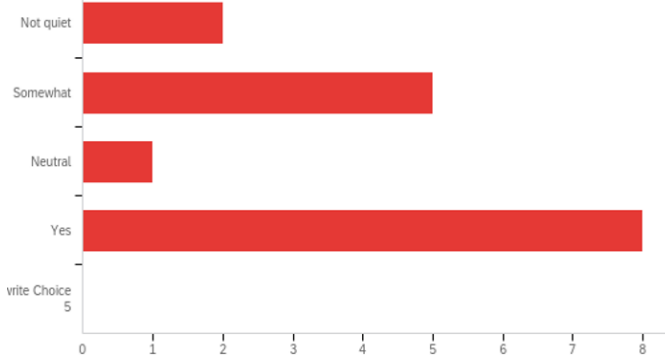

Fig. 5: Responses on question: Have you been ever educated on cybersecurity?

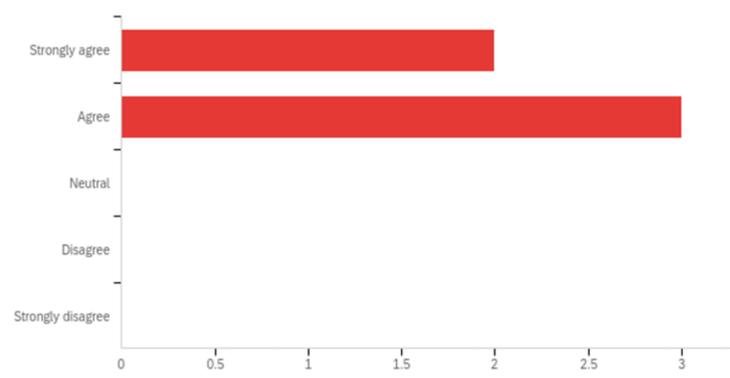

Fig. 6: Responses on: I learn better by personally doing or working through examples

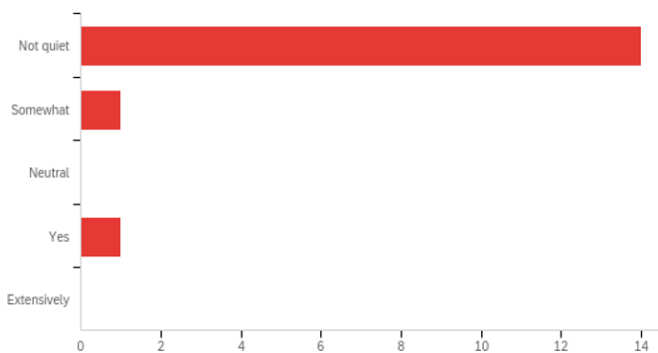

Fig. 4: Responses on question: Have you been ever educated on quantum machine learning?

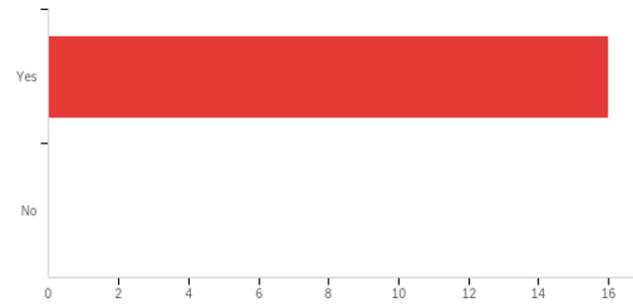

Fig. 7: Responses on: I agree to participate in this labware practice.

**Post-Test Survey**: We asked students if the tutorials in the pre-lab helped them understand more about the topics; in a post-test survey we completed it after involvement in the practical lab. Figure 9, 10, 11, and 12 displays the responses.

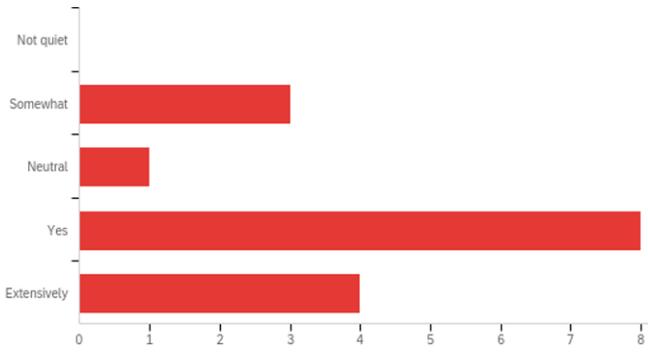

Fig. 8: Responses on: I like being able to work with the Case-based learning in Quantum Machine learning in cybersecurity.

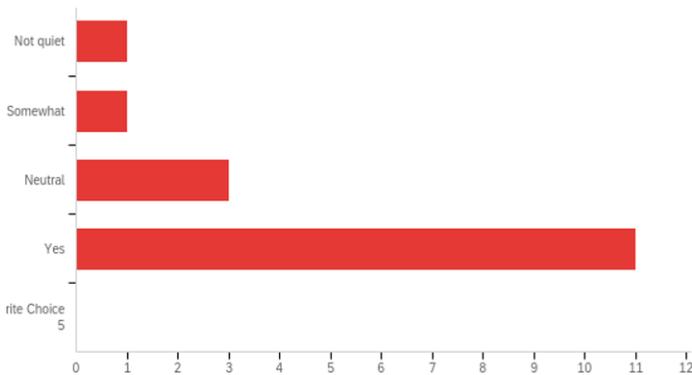

Fig. 9: Responses on: The outline of tutorials in the overview section help me learn more on the topics.

## VI. CONCLUSION

This labware seeks to address the demands and difficulties of learning with QML for cybersecurity through efficient and engaging case-study based learning techniques and the shortage of pedagogical resources and hands-on learning environments. The project offers a new teaching method for using QML to solve proactive cybersecurity issues. According to preliminary feedback, students learn from the concepts and practice the skills through the hands-on laboratories.

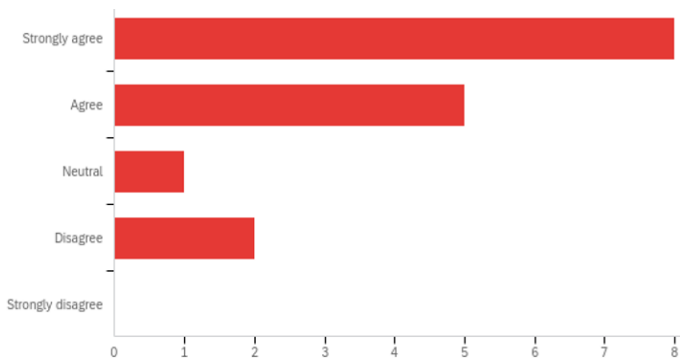

Fig. 10: Responses on: The hands-on prototype helps my learning experience on Quantum Machine learning in cybersecurity.

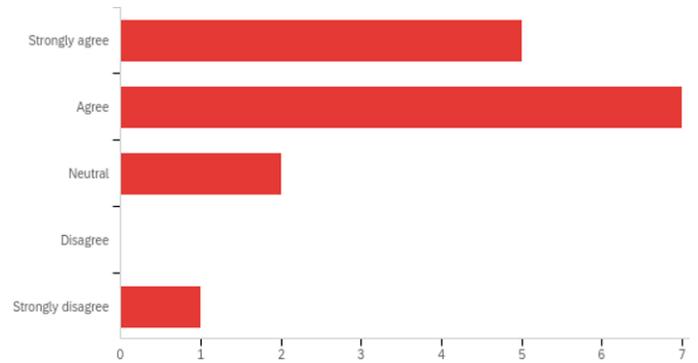

Fig. 11: Responses on: The case study-based learning modules help me apply learned knowledge to solve in Quantum Machine learning in cybersecurity problems in the future.

ACKNOWLEDGEMENT

The work is supported by the National Science Foundation under NSF Award #2100134, #2100115, #2209638, #2209637, #1663350, #2310179. Any opinions, findings, recommendations, expressed in this material are those of the authors and do not necessarily reflect the views of the National Science Foundation.